\documentclass[aps,pra,showpacs,amsmath,twocolumn,amssymb,nofootinbib,superscriptaddress]{revtex4-1}

\usepackage[pdftex]{graphicx}
\usepackage{mathrsfs,amsmath,amssymb,amsthm}
\usepackage[colorlinks=true,
            linkcolor=red,
            urlcolor=blue,
            citecolor=blue]{hyperref}

\begin{document}

\bibliographystyle{apsrev}

\title{Gaussian Beam-Propagation Theory for Nonlinear Optics ---\\
 Featuring an Exact Treatment of Orbital Angular Momentum Transfer}

\author{R. Nicholas Lanning}
\email[rlanni1@lsu.edu]{}
\author{Zhihao Xiao}
\affiliation{Hearne Institute for Theoretical Physics and Department of Physics $\&$ Astronomy, Louisiana State
University, Baton Rouge, Louisiana 70803, USA}
\author{Mi Zhang}
\author{Irina Novikova}
\author{Eugeniy E. Mikhailov}
\affiliation{Department of Physics, College of William $\&$ Mary, Williamsburg, Virginia 23187, USA}
\author{Jonathan P. Dowling}
\affiliation{Hearne Institute for Theoretical Physics and Department of Physics $\&$ Astronomy, Louisiana State
University, Baton Rouge, Louisiana 70803, USA}

\date{\today}

\begin{abstract}
We present a general, Gaussian spatial mode propagation formalism for describing the generation of higher order multi-spatial-mode beams generated during nonlinear interactions.
Furthermore, to implement the theory, we simulate optical angular momentum transfer interactions, and show how one can optimize the interaction to reduce the undesired modes. Past theoretical treatments of this problem have often been phenomenological, at best. Here we present an exact solution for the single-pass no-cavity regime, in which the the nonlinear interaction is not overly strong. We apply our theory to two experiments, with very good agreement, and give examples of several more configurations, easily tested in the laboratory. 
\end{abstract}
		

\maketitle

\section{Introduction}
\begin{figure}[b]
	\includegraphics[width=1.0\columnwidth]{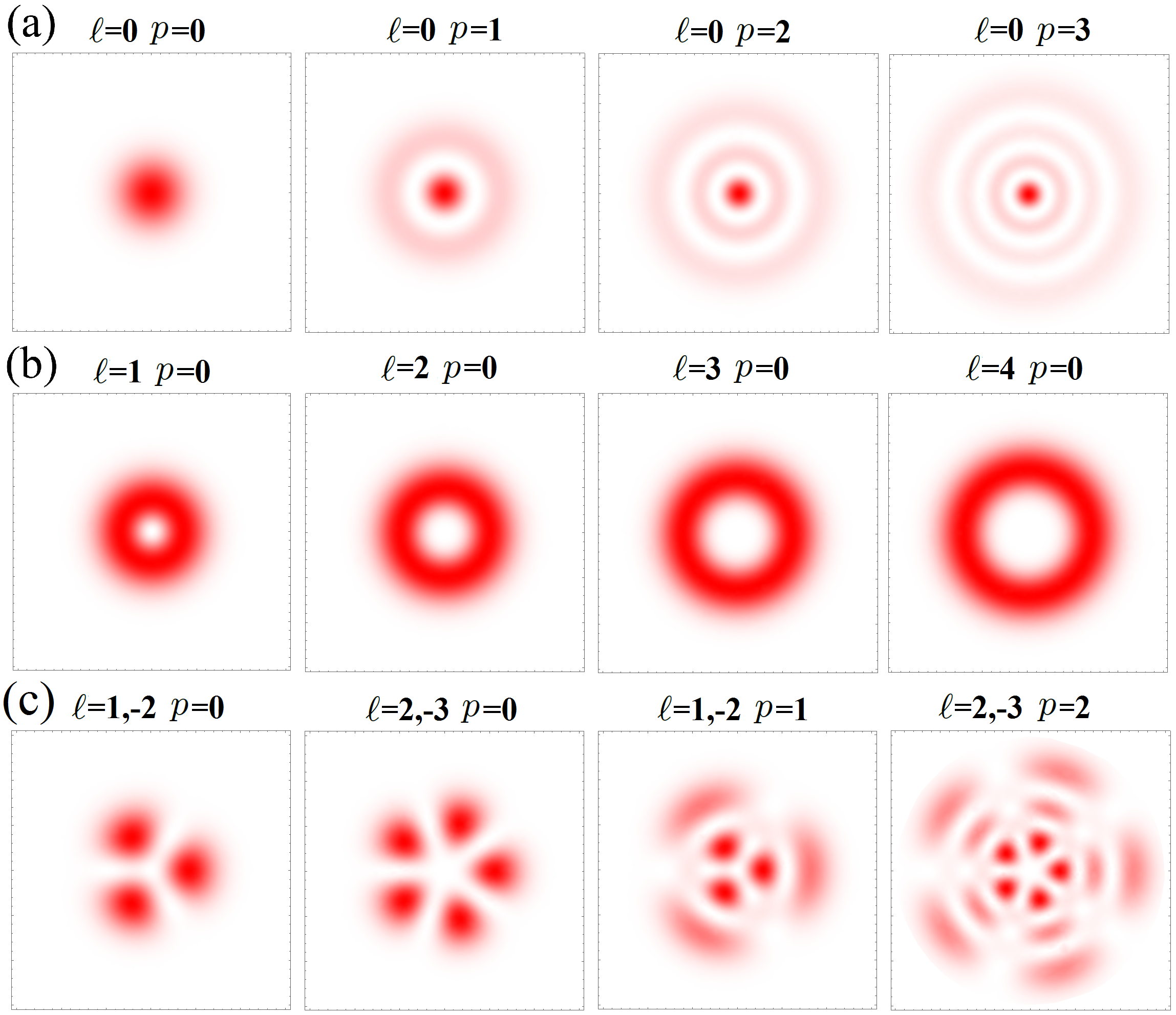}
	\caption{\label{fig:LG_EXAMPLES}LG mode intensity profiles for (a) $\ell =0$ and increasing $p$; (b) $p=0$ and increasing $\ell $; (c) for several superpositions of $\ell $ with $p=0,1,$ and $2$. Due to interference, superpositions with $\ell  \mathrm{'s}$ of opposite sign will have a symmetric petal-like structure with $|\ell _1|+|\ell _2|$ petals. See Ref. \cite{yao2011orbital} for an excellent description of optical OAM origins and behavior.
	}
\end{figure}

We are on the cusp of a new age of quantum physics and technology, where multi-spatial-mode beam propagation will play an ever more essential role.
Many of the most important quantum resources are produced during nonlinear light-matter interactions. 
It is particularly interesting to study nonlinear effects in response to beams carrying orbital angular momentum (OAM).
For example, the Laguerre-Gauss (LG) spatial modes have an azimuthal phase dependence of $\exp[i \ell \phi]$, which corresponds to OAM of $\ell \hbar$ per photon \cite{allen1992orbital, yao2011orbital}.
The intensity patterns for several LG modes, with $\ell \neq 0$, are given in Fig.~\ref{fig:LG_EXAMPLES}, and a more complete description can be found Sec. II C.
Conservation and storage (via slow and stopped light) of OAM has been realized in several processes including the entanglement of OAM modes in parametric down conversion (PDC) \cite{mair2001entanglement, caetano2002conservation}, second harmonic generation (SHG) \cite{courtial1997second}, and four-wave mixing (FWM) in semiconductors \cite{ueno2009coherent}.
Unlike solid state processes, nonlinear optics in atomic vapors is highly efficient and requires low-light intensities.
Transfer of OAM into atomic media \cite{pugatch2007topological, boyer2008entangled, cao2014transfer, de2015nonlinear}, transfer to frequency converted light \cite{tabosa1999optical, ding2012linear, walker2012trans, gariepy2014creating, zhdanova2015topological, akulshin2015distinguishing, akulshin2016arithmetic}, and amplification \cite{borba2016narrow} have all been observed in atomic vapors.

It has been observed that the transfer of OAM to frequency converted light in atomic media is typically accompanied by a disturbance of the radial component structure, i.e., frequency converted beams have nonzero radial index $p$, even when the pump beams have $p=0$ \cite{walker2012trans, zhang2016spatial, swaim2017atomic}. 
See Fig.~\ref{fig:LG_EXAMPLES} for examples of $p$ mode structure.
These effects may be subtle, or even negligible, when analysis is limited to comparing images collected with a charge-coupled device (CCD) camera. 
However, these effects can be quite detrimental to quantum processes such as squeezing \cite{zhang2016spatial} and entanglement generation \cite{dada2011experimental, romero2012increasing, fickler2012quantum, gupta2016multi}.
Nonlinear processes leading to the production of LG modes, with nonzero radial index $p$, can actually contaminate the mode structure and degrade the performance of the process.
As for the $\ell$ structure, the conservation of OAM dictates the allowed $\ell$ modes of the output beam, but this is not the end of the story. 
Transfer to some of these modes can be quite improbable \cite{walker2012trans, akulshin2015distinguishing}, even though the interaction would not violate OAM conservation. 
Therefore, an analysis simply based on conservation can be quite misleading.  
With these concerns in mind, we develop a relatively simple, yet analytic semiclassical theory, which predicts the spatial mode structure of beams created during nonlinear interactions.
We show that the predictions of our theory have excellent agreement with the observations we just discussed. 
Furthermore, the theory allows us to study and optimize the interaction to enhance performance and output mode structure.
Finally, an accurate description of the mode structure is a crucial ingredient in the general second-quantization procedure which we are currently developing.       
\section{Theoretical Background}
In the following section, we present an introduction to spatial mode propagation, properties of the Laguerre-Gauss modes, and an overview our method for solving the inhomogeneous wave equation in the paraxial approximation. 
An analytic method is preferred, because we wish to explicitly retain the mode structure of the resulting beam, such that one can use the mode structure, for example, to perform a second-quantization procedure. Please see the appendix for full details of our Green function solution method.
\subsection{Spatial Mode Propagation}
The paraxial wave equation and its solutions are certainly well known, but for completeness we quickly derive a form useful for our purposes. 
We investigate the electric field propagating through a nonlinear medium, which for simplicity, we treat as lossless and dispersionless. 
Treating the light classically, we start with the inhomogeneous wave equation
\begin{equation}\label{eq:INHOMO_WAVE_EQ}
\nabla^{2}\tilde{\textbf{E}}-\frac{1}{c^{2}}\frac{\partial^{2} \tilde{\textbf{E}}}{\partial t^{2}}=	\frac{1}{\epsilon_{0} c^{2}} \frac{\partial^{2}\tilde{\textbf{P}}}{\partial t^{2}},
 \end{equation}
where the tilde indicates rapidly varying quantities. 
Introducing the field $\tilde{\textbf{E}}(\textbf{r},t)=\textbf{E}(\textbf{r})e^{-i \omega t}$ and polarization $\tilde{\textbf{P}}(\textbf{r},t)=\textbf{P}(\textbf{r})e^{-i \omega t}$ into Eq.~(\ref{eq:INHOMO_WAVE_EQ}), one finds the Helmholtz equation.
Next, we let $\textbf{E}(\textbf{r}) = \boldsymbol{\mathcal{E}}(\textbf{r}_{\perp},z) e^{ikz}$, $\textbf{P}(\textbf{r}) = \boldsymbol{\mathcal{P}}(\textbf{r}_{\perp},z) e^{ikz}$, and along with the paraxial approximation ($|k \, \partial \mathcal{E} / \partial z| \gg |\partial^2 \mathcal{E} / \partial	z^2|$) we transform the Helmholtz equation into the inhomogeneous paraxial wave equation
 \begin{equation}\label{eq:PARAXIAL_WAVE_EQ}
(\frac{\partial}{\partial z}-\frac{i}{2k}\nabla^{2}_{\perp})\boldsymbol{\mathcal{E}}=\frac{i k}{ 2 \epsilon_{0}} \boldsymbol{\mathcal{P}}.
 \end{equation}
Depending on the scenario, there are several methods used for deriving the polarization $\boldsymbol{\mathcal{P}}$, and in many cases our simple phenomenological approach will lead to interesting and accurate results. 
Regardless of the method used, the polarization is, in general, a complicated function of the fields, that is, $\boldsymbol{\mathcal{P}}=\boldsymbol{\mathcal{P}}(\boldsymbol{\mathcal{E}})$.
This makes Eq. (\ref{eq:PARAXIAL_WAVE_EQ}) similar in structure to the nonlinear Schr$\ddot{\mathrm{o}}$dinger equation.
With this similarity in mind, we make a first order Born approximation \cite{saleh1991fundamentals}, and replace $\boldsymbol{\mathcal{E}}$ with the input beam $\boldsymbol{\mathcal{E}}_0$ instead, which establishes a much simpler differential equation of the form,
  \begin{equation}\label{eq:DE1}
(\frac{\partial}{\partial z}-\frac{i}{2k}\nabla^{2}_{\perp})\boldsymbol{\mathcal{E}}=\frac{i k}{ 2 \epsilon_{0}} \boldsymbol{\mathcal{P}}(\boldsymbol{\mathcal{E}}_0) \equiv \boldsymbol{\wp}(\boldsymbol{\mathcal{E}}_0).
 \end{equation}
The first born approximation is sufficient when the pump beam is only slightly modified and the seed or signal fields are weak compared to the driving fields.
Fortunately, most nonlinear processes that involve generation of new fields start with vacuum modes.
Furthermore, the nonlinear processes that we consider in this article satisfy these conditions.
For example, in the polarization self-rotation squeezing scheme \cite{matsko2002vacuum, ries2003experimental, zhang2013generating}, the seed beam is just vacuum entering into an empty port.
Thus, in the Born approximation, the optical response of the nonlinear material inherent, in $\boldsymbol{\wp}(\boldsymbol{\mathcal{E}}_0)$, behaves as a source for new mode components of the field.
Regarding the vector nature of the interaction, it must must be noted that in general, the nonlinear susceptibility governing the interaction has a tensor form with, e.g., 81 terms for third-order nonlinearities.
This structure offers a complication, which can lead to a set of coupled equations that are very difficult to solve.
Fortunately, the majority of nonlinear processes that involve generation of new fields have symmetry, which vastly reduces the complication of the problem.
This fact, paired with the Born approximation, is used to simplify the problem to uncoupled differential equations, which then can be solved using our method.   
\subsection{Construction of the initial-value-problem}
In the first Born approximation, the right-hand-side of Eq.~(\ref{eq:DE1}) is effectively the source of the new beam. Thus, it is helpful to restate the problem here as an initial-value problem (IVP) in the compact form,
\begin{equation} \label{eq:DE2}
\begin{split}
\mathrm{DE:}&\;\; \hat{L}\mathcal{E} = \wp(\mathcal{E}_0)\\
\mathrm{IV:}&\;\;\;\;\;\;\;\; \mathcal{E}_{0},
\end{split} 
\end{equation}
where $\hat{L} \equiv \frac{\partial}{\partial z}-\frac{i}{2k}\nabla^{2}_{\perp}$ and $\mathcal{E}_{0}$ is the input-field amplitude which is assumed to be undepleted.
Although this IVP represents a first Born approximation, the following method can be used iteratively for cases when the nonlinearity and generated fields are stronger. 
In such cases, it may be necessary to include an attenuation factor, on the initial value, which conserves energy.   
Ideally, the theory does not require any free parameters for the source $\wp$, and can be used directly. 
In reality, however, it may be difficult to calculate the strength of the source if the model used does not capture the full complexity of the experimental system.
Therefore, it maybe convenient to capture such complexity with an interaction strength factor that, if necessary, can be used as a fitting parameter.
For example, a fitting parameter is quite useful when simulating the generation of squeezed light in hot atomic vapor, where the direct calculations are too complex~\cite{zhang2016spatial}.
\subsection{Laguerre-Gauss Mode Properties}
We elect to consider input beams with cylindrical symmetry, but we note that the following calculation can certainly be done in other coordinate systems. 
In cylindrical coordinates, the homogeneous paraxial wave equation gives rise to the LG family of solutions \cite{Siegman_book}:
\begin{equation} \label{eq:LG_MODES}
 \begin{split}
  	u_{\ell ,p}(\vec{r})=& \dfrac{C_{\ell ,p}}{w(z)} e^{- \frac{r^{2}}{w(z)^{2}}}e^ { -\frac{ikr^{2}z}{2(z^{2}+z_{R}^{2})} }  \big( \dfrac{\sqrt{2}r}{w(z)} \big)^{|\ell |} \\
	\times&  \: L_{p}^{|\ell |}
	\big( \dfrac{2r^{2}}{w(z)^{2}} \big)  e^{i \ell \phi} e^{i(2p+|\ell |+1)\arctan(z/z_{R})}, 
 \end{split}
\end{equation}
where $\ell$ is the azimuthal index, $p$ is the radial index for each mode, $ C_{\ell ,p}=\sqrt{2p! / \pi(|\ell |+p)!}$ is a normalization constant, $w_{0}$ is the beam waist, $ w(z)=w_{0}\sqrt{1+(z/z_{R})^2}$ is the width function of the beam, $L_{p}^{|\ell |}$ are the generalized Laguerre polynomials, $z_{R}=\pi w_{0}^{2}/\lambda$ is the Rayleigh range, and $k=2\pi/\lambda$ is the wave
number. 
The LG modes form a complete orthonormal set and thus can be used as a basis set to expand an arbitrary paraxial beam $B=B(r,\phi,z)$ in free space. Using the orthogonality relation
\begin{equation} \label{eq:ORTHOG}
\int rdrd\phi \, u_{\ell ,p}^{*}(r,\phi,z)u_{q,n}(r,\phi,z)=\delta_{\ell q}\delta_{pn},
\end{equation}
we can write $B(r,\phi,z)$ as
\begin{equation} \label{eq:B_EXPAN}
\begin{split}
B(r,\phi,z) = \sum_{\ell ,p} c_{\ell ,p}(w_0) u_{\ell ,p}(r,\phi,z,w_0),
\end{split}
\end{equation}
where
\begin{equation} \label{eq:B_EXPAN_COEFF}
c_{\ell ,p}(w_0) = \int r dr d\phi \, u_{\ell ,p}^{*}(r,\phi,z_0, w_0) B(r,\phi,z_0).
\end{equation}
The waist $w_0$ of the basis set is, in general, chosen to give the best fit and reduce the number of terms in the expansion, whereas the $c_{\ell ,p}$ coefficients are independent of the position $z_0$. 
If we insert Eq.~(\ref{eq:B_EXPAN_COEFF}) in Eq.~(\ref{eq:B_EXPAN}) and collect the LG modes we find 
\begin{equation}
\begin{split}
B(r,\phi,z) =& \int r' dr' d\phi' \, B(r',\phi',z_{0})\\ 
*& \Big( \sum_{l,p} \, u_{l,p}^{*}(r',\phi',z_{0})   u_{l,p}(r,\phi,z) \Big).
\end{split}
\end{equation} 
If we impose $z = z_{0}$, then we can establish the very important completeness relation:
\begin{equation} \label{eq:COMPLETE}
\sum_{\ell ,p}\,u_{\ell ,p}^{*}(r',\phi',z)u_{\ell ,p}(r,\phi,z)=\delta(r-r')\delta(\phi-\phi').
\end{equation}
The condition $z = z_{0}$ simply states that the LG modes are complete at equal $z$'s, i.e, when the $z$-slices chosen coincide. 
Furthermore, we will see that this condition reemerges as part of the mechanism which introduces the input beam in the solution of the IVP.  
Thus, the completeness relation is instrumental in the Green function solution method, which we will now discuss briefly.  
\subsection{Green Function Solution}
The magic of the Green function solution method \citep{Barton_book} is that once the propagator $K$, and Green function $G$ are derived, the problem is solved (for full details of the solution method see the appendix):
\begin{equation}\label{eq:MAGIC_RULE}
\begin{split}
\mathcal{E}& =  \int r'dr'd\phi'\,K(\textbf{r} \, | \, \textbf{r} ')\mathcal{E}_{0}(\textbf{r} ') \; |_{z=z'}\\
&+ \int dz' \int r'dr'd\phi' \, G(\textbf{r} \, | \, \textbf{r} ')\wp (\textbf{r} ').
\end{split}
\end{equation}
Although we solve this problem in free space, the properties of the LG modes allow us to utilize a method developed for fixed boundary conditions.
Since the LG modes are a complete orthonormal set, we can use an eigenfunction expansion.
Recalling the completeness relation in Eq.~(\ref{eq:COMPLETE}), we define our propagator and Green function in terms of the LG modes:
\begin{equation}\label{eq:KANDG}
\begin{split}
K(r,\phi,z|r',\phi',z') &\equiv \sum_{\ell ,p}\,u_{\ell ,p}^{*}(r',\phi',z')u_{\ell ,p}(r,\phi,z)\\
G(r,\phi,z|r',\phi',z') &\equiv \Theta(z-z') K(r,\phi,z|r',\phi',z'),
\end{split}
\end{equation}
where $\Theta(z-z')$ is the Heaviside step function.

At this point the problem is solved.
However, we may further simplify matters by expanding the source $\wp$ in terms of the LG modes: 
\begin{equation}\label{eq:SOURCE}
\wp(r,\phi,z)=\sum_{l,p}c_{l,p}(z) \, u_{l,p}(r,\phi,z),
\end{equation}
where $c_{l,p}(z)=\int r dr d\phi\,u_{l,p}^*(r,\phi,z)\wp(r,\phi,z)$.
To find the final form of our solution, we insert Eq.~(\ref{eq:KANDG}) and Eq.~(\ref{eq:SOURCE}) into Eq.~(\ref{eq:MAGIC_RULE}) and find
\begin{equation}\label{eq:FINAL_SOLN2}
\mathcal{E}(\textbf{r})=\mathcal{E}_0(\textbf{r}) + \sum_{l,p}u_{l,p}(\textbf{r})\int_{z_i}^{z_f} dz' c_{l,p}(z').
\end{equation}
So we see that the final solution is a superposition of the unmodified pump beam, and a collection of LG modes, which, referring to Eq.~(\ref{eq:DE1}), will depend on the specified polarization $P$ and the spatial structure of the input beam $\mathcal{E}_0$. The cautious reader may be concerned that we have solved a problem concerning a possibly localized source distribution using a free space Green function method. We pause to point out that the effects of the source are totally subsumed in the spacial distribution of $\wp$. Furthermore, in most situations a Gaussian pump beam will be completely encompassed by the interaction region, effectively making the boundary in $r$ infinite.
\pagebreak
\section{Simulation of Experiment}
We start will several simulations which showcase our theory.
We first focus on experimentally relevant simulations related to PDC and spontaneous FWM.  
For PDC, we take the results of the experiment a step further by suggesting ways to enhance the mode structure of the generated beam.
For FWM, we address an unresolved problem pertaining to the pathways of OAM transfer in spontaneous FWM.
For each simulation, we will specify the polarization governing the interaction, along with the relevant beam and material parameters.
Furthermore, all beam profiles are plotted at $z=0$, and are color coded to match the visible spectrum as closely as possible.
\subsection{Stimulated Down Conversion}
\begin{figure}[b]
	\includegraphics[width=1.0\columnwidth]{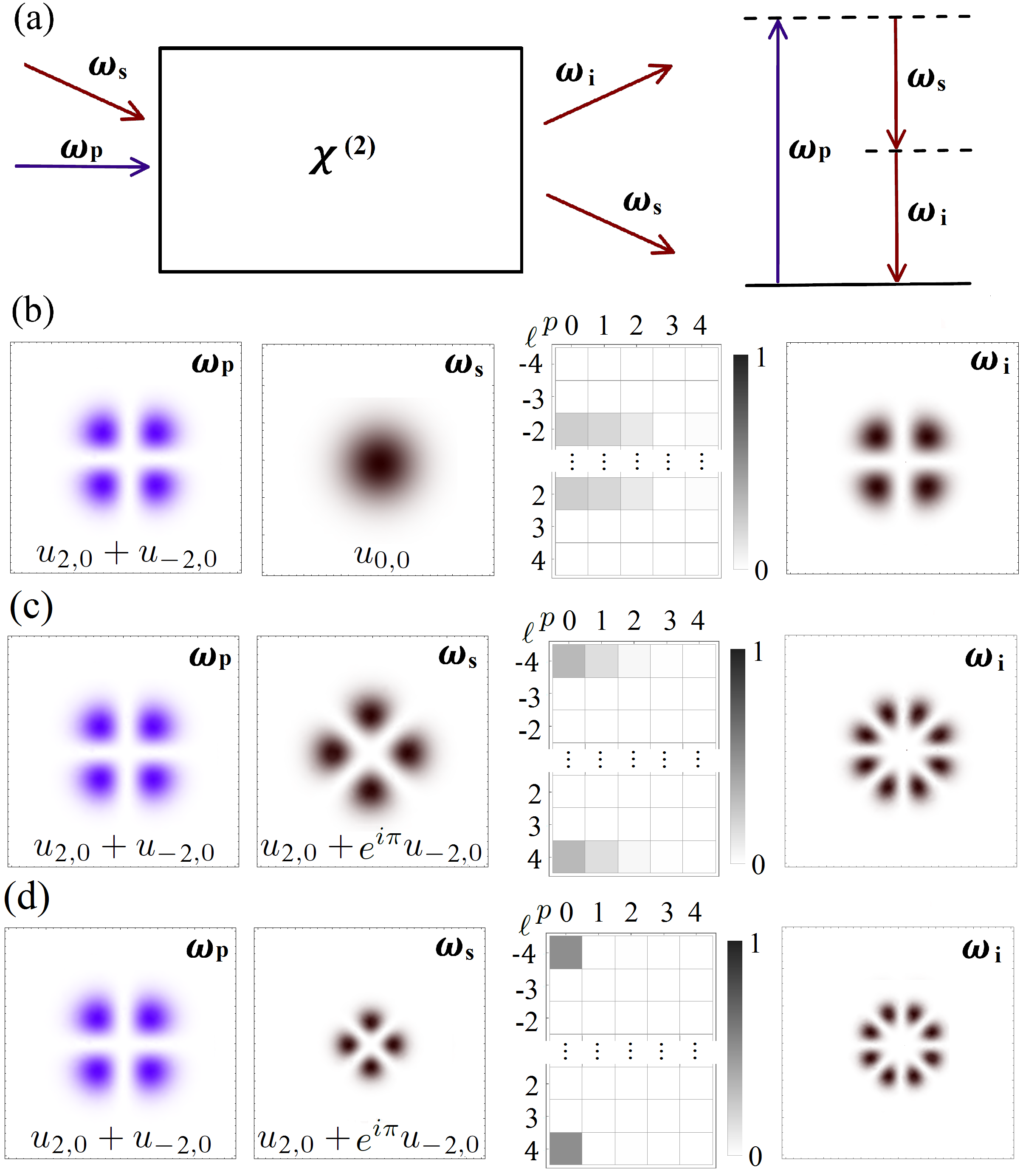}
	\caption{\label{fig:DOWNCONV}In (a) we depict the simple stimulated-down-conversion scheme of our simulation. In (b) through (c) we show the input beam profiles of the pump and signal beams, a histogram giving the mode structure of the output (idler) beam, and the spatial profile of the output (idler) beam respectively. In (b) we use a $u_{0,0}$ signal as a baseline, then in (c) and (d), we show how the signal beam can be chosen to tailor a clean higher OAM superposition in the idler mode. 
	}
\end{figure}

First, we will investigate OAM transfer in frequency down-converted light. 
A rapidly evolving body of theoretical work is calling for entanglement in quantum systems with higher dimensions \cite{groblacher2006experimental, mirhosseini2015high}. 
Thus, as an exercise, we will study stimulated PDC as a way to understand the properties of down-converted twin beams when a cavity is not present to alter the mode structure.
We model a PDC experiment, in which a $442 \, \mathrm{nm}$ pump beam interacts with a $845 \, \mathrm{nm}$ signal beam (in a $3 \, \mathrm{mm}$ long BBO crystal) to create a $925 \, \mathrm{nm}$ idler beam \cite{caetano2002conservation}. 
Due to the small incident angle of the signal beam and the relatively thin crystal, effects from the non-collinear geometry are negligible (see appendix). 
The light-matter interaction in the crystal is governed by the polarization $\mathcal{P}(\omega_i)^{(2)} =\epsilon_0 \chi^{(2)} \mathcal{E}_p(\omega_p) \mathcal{E}_s(\omega_s)^*$, depicted in Fig.~\ref{fig:DOWNCONV}(a).
In this interaction $E_p$ is the pump beam, $E_s$ is the signal beam, and the conjugation of $E_s$ corresponds to the creation of a photon in the input-signal mode. 
Therefore, we see how the choice of pump and signal modes tailor the response of the idler beam; this phenomenon has been verified experimentally for relatively simple input beams \cite{caetano2002conservation}.
The waist of the basis set, used to represent the generated beam, is set to match the waist of the pump beam.
In this simulation, we will see how the interplay of the waist and Rayleigh range, of the input beams, effects the mode structure of the generated beam. 

In this simulation, we investigate how to generate a high-OAM superposition in the idler mode. 
First, for reference, we consider that the signal beam is prepared in the $u_{0,0}$ mode with a waist that matches the pump beam. 
As one can see in Fig.~\ref{fig:DOWNCONV}(b), the idler beam responds in $\ell =\pm2$ superposition, with a spreading in $p$ modes, corresponding to the partial overlap of the modes in $E_p(\omega_p) E_s(\omega_s)^*$.
Next, we investigate how to increase the OAM in the idler beam. We prepare a signal beam in the same superposition as the pump, but include a $\pi$ phase shift between the two modes in the superposition, i.e., the signal beam is now $u_{2,0} + e^{i \pi} u_{-2,0}$. 
In Fig.~\ref{fig:DOWNCONV}(c) we see that this combination creates a destructive interference, which suppresses response at $\ell =0$, and produces an idler beam, which responds at $\ell =\pm4$ with a spreading in $p$ modes. More simulations using this technique can be found in the following section. 
Again, the spreading in $p$ modes corresponds, in part, to the poor overlap of the beams. 
Therefore, in Fig.~\ref{fig:DOWNCONV}(d), we show that one can optimize the signal-beam waist to suppress the higher order $p$ modes. 
The narrowing of the signal beam waist $w_0$ (by $\approx \, 40 \% $), in Fig.~\ref{fig:DOWNCONV}(d), corresponds to a reduction of the Rayleigh range $z_R \equiv \pi w_{0}^{2} / \lambda$. 
In effect, the beams now expand closer to the same rate, which is apparently the optimized beam overlap through the interaction region. 
This observation is related to the Boyd criterion \citep{walker2012trans}, which states that, the nonlinear interaction is strongest when the Rayleigh ranges of the interacting beams are identical. 
The effect observed here extends this observation to include an improvement in the mode structure. 
This simulation is very useful, since one cannot determine the $p$ mode structure experimentally by simply analyzing the intensity pattern, e.g., differences in the petal structures of $\omega_i$, in Fig.~\ref{fig:DOWNCONV}(c,d), are not discernible, even though (c) has a contaminated mode structure.        
\subsection{Spontaneous Four-Wave Mixing}
\begin{figure}[t]
	\includegraphics[width=1.0\columnwidth]{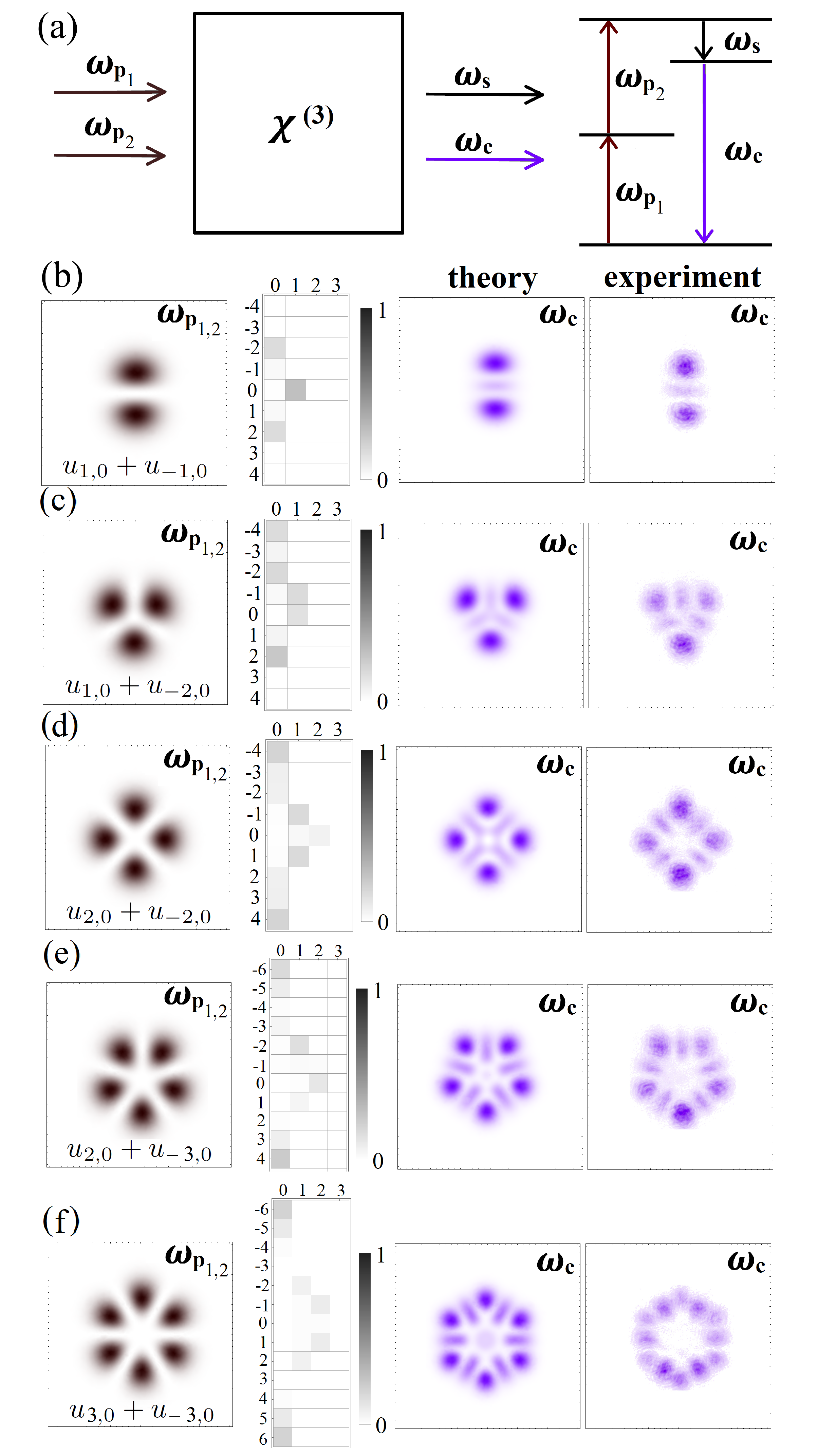}
	\caption{\label{fig:FWM2}In (a) we depict a non-degenerate four-wave mixing scheme, which occurs in Rb87. In (b) through (f), we study the result of this interaction for different pump profiles; from left to right the columns correspond to the profile of the pump beams, a histogram giving the mode structure of the conjugate beam, the spatial profile of the conjugate beam, and lastly an experimental realization from Ref. \citep{walker2012trans}. A better qualitative agreement is observed when $\omega_s$ is allowed to take on any mode conserving OAM. This result suggests that, contrary to previous assumptions, that the long wavelength signal beam can in fact carry OAM.
	}
\end{figure}

Next, we present a simulation for non-degenerate four wave mixing. 
In this scheme, two pump beams spontaneously create signal and conjugate beams, according to the polarization $\mathcal{P}(\omega_c)^{(3)}=\epsilon_{0} \chi^{(3)} \mathcal{E}(\omega_{p1})\mathcal{E}(\omega_{p2})\mathcal{E}(\omega_s)^*$; depicted in Fig.~\ref{fig:FWM2}(a). 
In our other simulations, all of the fields which generate the new beam were carefully selected input beams. 
This last simulation is distinct since the signal and conjugate beams are spontaneously created during the interaction. 
Therefore, to predict the mode structure of the conjugate beam, one must make assumptions about the spontaneous response creating the signal beam. 
The question remains, how do we predict the coupling of the atoms to the vacuum LG modes at the signal and conjugate wavelengths?

To investigate this problem, we simulate an interaction, which has been experimentally realized \citep{walker2012trans}. A $780 \, \mathrm{nm}$ pump and a $776 \, \mathrm{nm}$ pump stimulate a nonlinear interaction in Rb, which is the source of a $5230 \, \mathrm{nm}$ signal beam and a $420 \, \mathrm{nm}$ conjugate beam [see Fig.~\ref{fig:FWM2}(a)].
In the experiment, the long wavelength beam remains unobserved. 
However unlikely, OAM transfer to this beam should not be ruled out as a possibility. 
Rather, the atoms can couple to any spatial light mode, which obeys OAM conservation, and thus we should account for each of these modes in our simulation.
Therefore, for the sake of simulation, we input a signal beam that is a balanced incoherent superposition of all OAM modes, which obey OAM conservation, and allow the integral in Eq.~(\ref{eq:FINAL_SOLN2}) to determine which modes are populated. 
Thus, we accommodate for the possible OAM transfer to the long-wavelength beam, and only restrict our analysis to incoherent superposition at the signal wavelength. 
The last assumption we will impose for this simulation is the matching of the Rayleigh ranges for all beams, i.e., we invoke the Boyd criterion \citep{walker2012trans}.
In effect, we judiciously choose the widths of the beams to maximize the nonliniear interaction.

The results are presented in Fig.~\ref{fig:FWM2}(b--f). 
From left to right, the columns correspond to the profile of the pumps, a histogram giving the mode structure of the conjugate beam, the profile of the conjugate beam, and lastly the experimental data from Ref. \citep{walker2012trans}. 
We see excellent qualitative agreement, when the signal beam is allowed to carry OAM. 
In particular, our simulation appears to take account of the relative brightness and shape of the lobes in the experimental data. 
In each case there are lobes that are elongated and dimmer than others and, our simulation agrees with this observation. 
It is apparent that the richer mode structure that we predict for the conjugate beam can account for the variations in the lobe brightness, without drastically affecting the lobe structure. 
Therefore, we emphasize that a more careful analysis, such as the one we present here, is necessary when studying the mode structure of output beams, since a naive examination of the lobe structure can be very misleading.
\section{Predictive Simulations}
In the previous examples, we have shown how our theory predicts the mode structure during OAM transfer.
Now we will show how it can be used as a tool to optimize the mode structure for OAM addition, subtraction, and cancellation processes.
The theme of these simulations is to develop methods which enhance the mode structure of the generated beams. 
They also serve as predictions which can be verified experimentally, relatively easily. 
\subsection{Second Harmonic Generation}
\begin{figure}[b]
	\includegraphics[width=1.0\columnwidth]{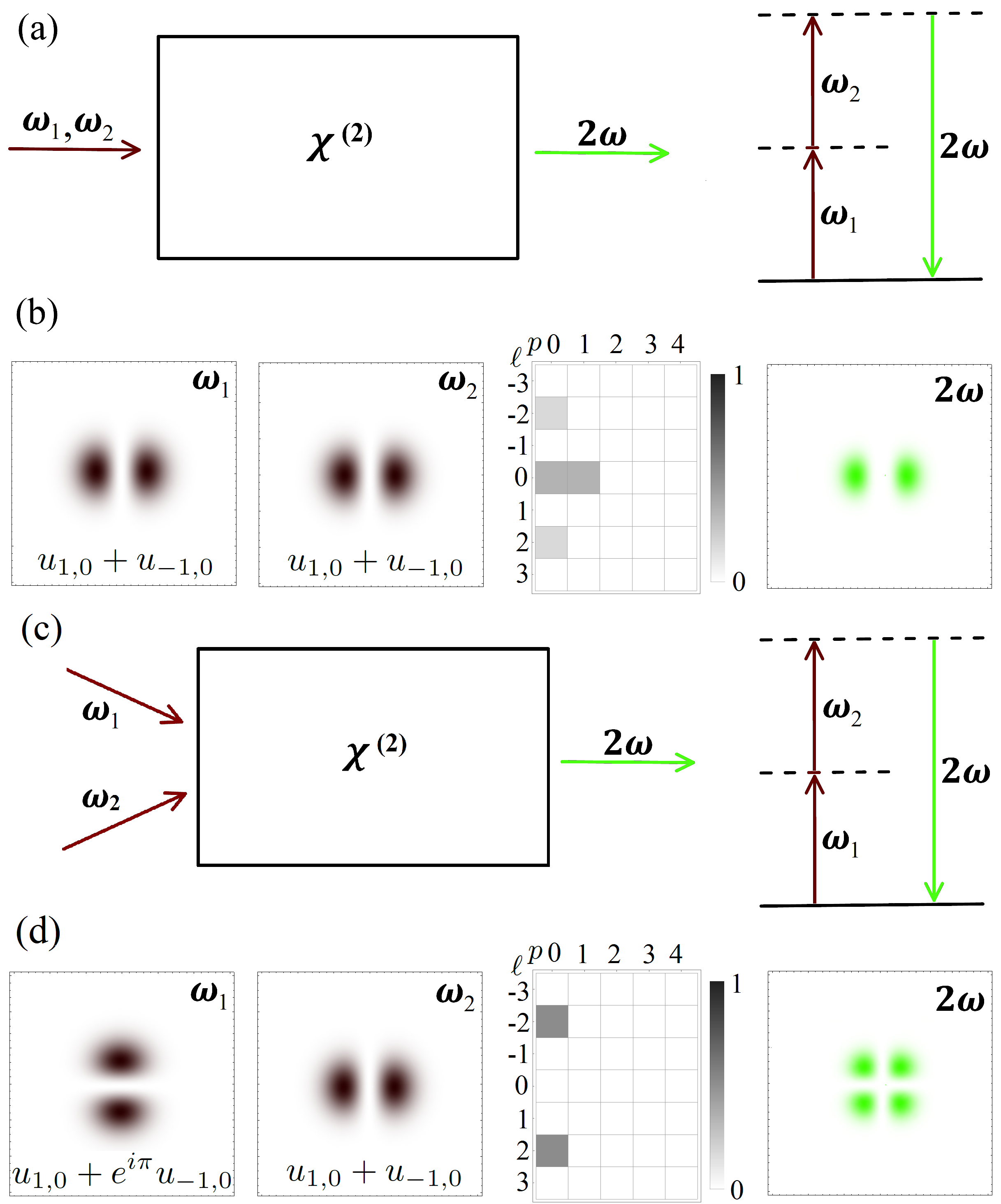}
	\caption{\label{fig:2ndHARM}In (a) and (c) we depict the two simple second-harmonic generation schemes of our simulation. In (b) and (d) we show the input-beam profile, which is supplying the two pump photons, a histogram giving the mode structure of the output beam, and the spatial profile of the output beam, respectively. In (b) both pump photons come from a $u_{1,0}+u_{-1,0}$ superposition. In (c) and (d), a phase shift and a new geometry is chosen such that destructive interference cancels the response at $\ell =0$ [seen in (b)]. The pump photons are in the far-infrared at 1140 nm, and thus the 2nd harmonic response is at 570 nm. 
	}
\end{figure}

First we present a simulation for second harmonic generation.
An $1140 \, \mathrm{nm}$ pump beam, with waist $w_0 = 0.1$ mm, interacts with a $3 \, \mathrm{mm}$ long crystal to create a second harmonic beam at $570 \, \mathrm{nm}$. 
The waist of the basis set, used to represent the generated beam, is set to match the waist of the pump beams.
The light-matter interaction in the crystal is governed by the polarization $\mathcal{P}(2\omega)^{(2)}=\epsilon_{0} \chi^{(2)} \mathcal{E}(\omega_1)\mathcal{E}(\omega_2)$, depicted in Fig.~\ref{fig:2ndHARM}(a).

We first consider that two photons are annihilated from a single pump beam in the superposition $u_{1,0}+u_{-1,0}$, and the results are given in Fig.~\ref{fig:2ndHARM}(b).
Each histogram depicts the probabilities $
\mathscr{P}_{\ell ,p}$ of modes being excited, i.e., referring to Eq.~(\ref{eq:FINAL_SOLN2}), $\mathscr{P}_{\ell ,p} \equiv |\int dz' c_{\ell ,p}(z')|^2 / \sum_{\ell,p}|c_{\ell,p}|^2$.
One can see that along with the response at $\ell =\pm2$, there is also response at $\ell =0$, corresponding the cross terms in $(u_{1,0}+u_{-1,0})^2$. 
One can enhance the OAM transfer by creating a destructive interference to remove the response at $\ell =0$.
This can be done by including a second pump beam, according to the geometry in Fig.~\ref{fig:2ndHARM}(c), with a $\pi$ phase shift in the superposition, i.e., a rotated profile. 
The conservation of linear momentum dictates that, in order to have a response at $2\omega$ in the geometry of Fig.~\ref{fig:2ndHARM}(c), a photon must be annihilated from each pump beam. 
As one can see in Fig.~\ref{fig:2ndHARM}(d), this geometry (along with the phase shift) suppresses the response at $\ell =0$ and provides a cleaner mode structure. 
This technique can be quite useful and can be implemented in more complex situations, as we will show in the following examples.
\subsection{Third Harmonic Generation}
\begin{figure}[t]
	\includegraphics[width=1.0\columnwidth]{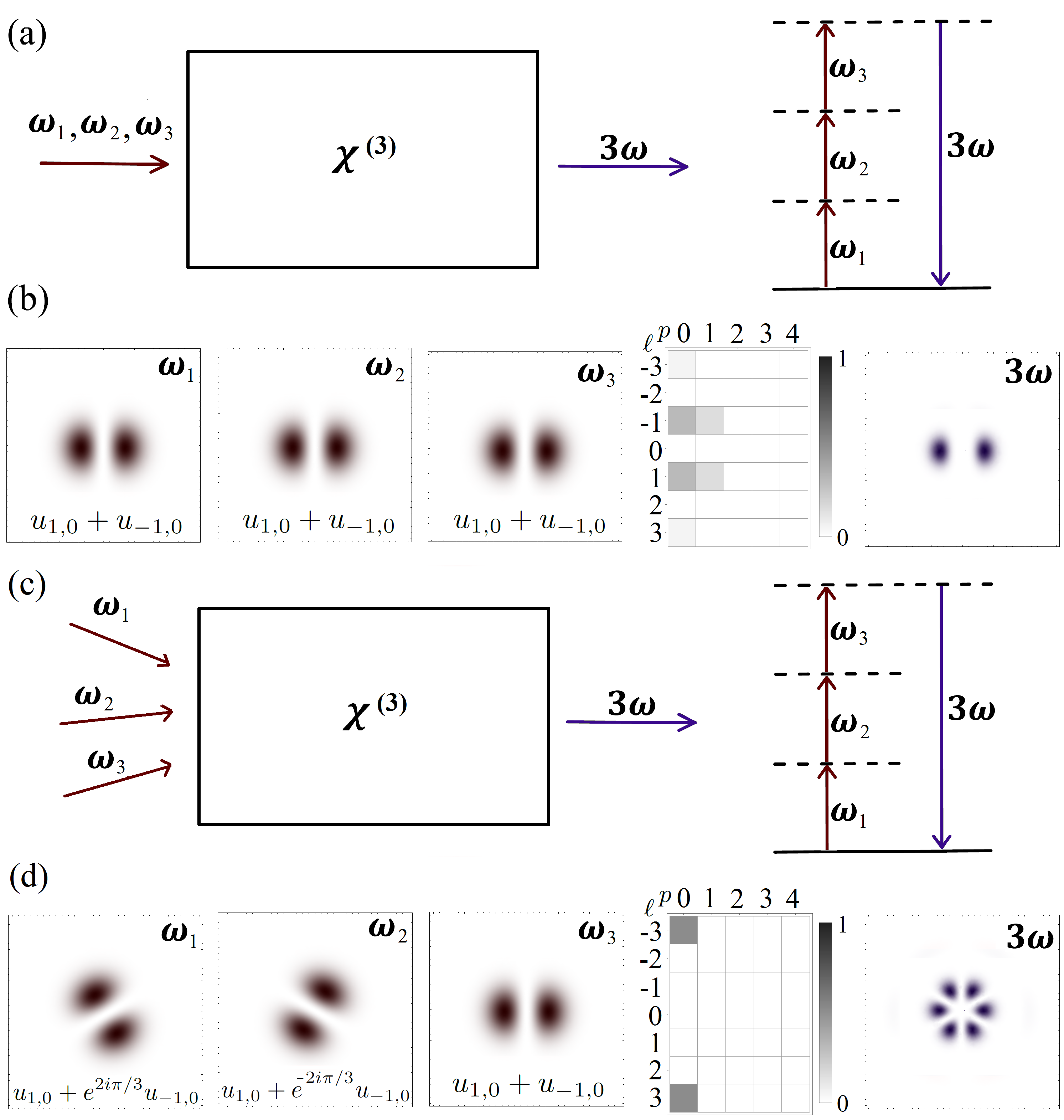}
	\caption{\label{fig:3rdHARM}In (a) and (c) we depict the two simple third-harmonic generation schemes of our simulation. In (b) and (d) we show the input-beam profiles, which are supplying the three pump photons, a histogram giving the mode structure of the output beam, and the spatial profile of the output beam, respectively. In (b) all pump photons come from a $u_{1,0}+u_{-1,0}$ superposition. In (c) and (d), phase shifts and a new geometry is chosen such that destructive interference cancels the response at $\ell =\pm 1$ as seen in (b). The pump photons are in the far-infrared at 1140 nm, and thus the 3rd harmonic response is at 380 nm.
	}
\end{figure}

Next, we present a simulation for third-harmonic generation. An $1140 \, \mathrm{nm}$ pump beam, with waist $w_0 = 0.1 \, \mathrm{mm}$, interacts with a $3 \, \mathrm{mm}$ long crystal to create a third-harmonic beam at $380 \, \mathrm{nm}$. The light-matter interaction in the crystal is governed by the polarization $\mathcal{P}(3\omega)^{(2)}=\epsilon_{0} \chi^{(2)} \mathcal{E}(\omega_1)\mathcal{E}(\omega_2)\mathcal{E}(\omega_3)$, depicted in Fig.~\ref{fig:3rdHARM}(a).
As in the previous simulation, we first consider that three pump photons are annihilated from a single pump beam in the superposition $u_{1,0}+u_{-1,0}$, and the results are given in Fig.~\ref{fig:3rdHARM}(b). 
One can see from the histogram that, along with the response at $\ell =\pm3$, there is also response at $\ell =\pm1$, corresponding the cross terms in $(u_{1,0}+u_{-1,0})^3$. 
To clean up the OAM transfer, one needs to create an interference to destroy the response at $\ell =\pm 1$. 
We deploy the same tactic as in the SHG simulation, but in this case include three pump beams, two of which have the phase shift $\pm 2\pi /3$ between the two modes. 
One can see in Fig.~\ref{fig:3rdHARM}(c,d) that if one photon is annihilated from each of the three pump beams, that the response at $\ell =\pm1$ is destroyed and a clean OAM transfer is established.
Again, as we described in the previous subsection, one could realize this result by judiciously choosing the angles of the incident beams according to conservation of linear momentum, such that the generated beam is aligned along the $z\mathrm{-axis}$. 
So we see that this can be a very useful technique for the up-conversion of OAM, especially in the absence of a cavity, which would naturally clean the mode structure.        
\subsection{Four-wave Mixing}
\begin{figure}[t]
	\includegraphics[width=1.0\columnwidth]{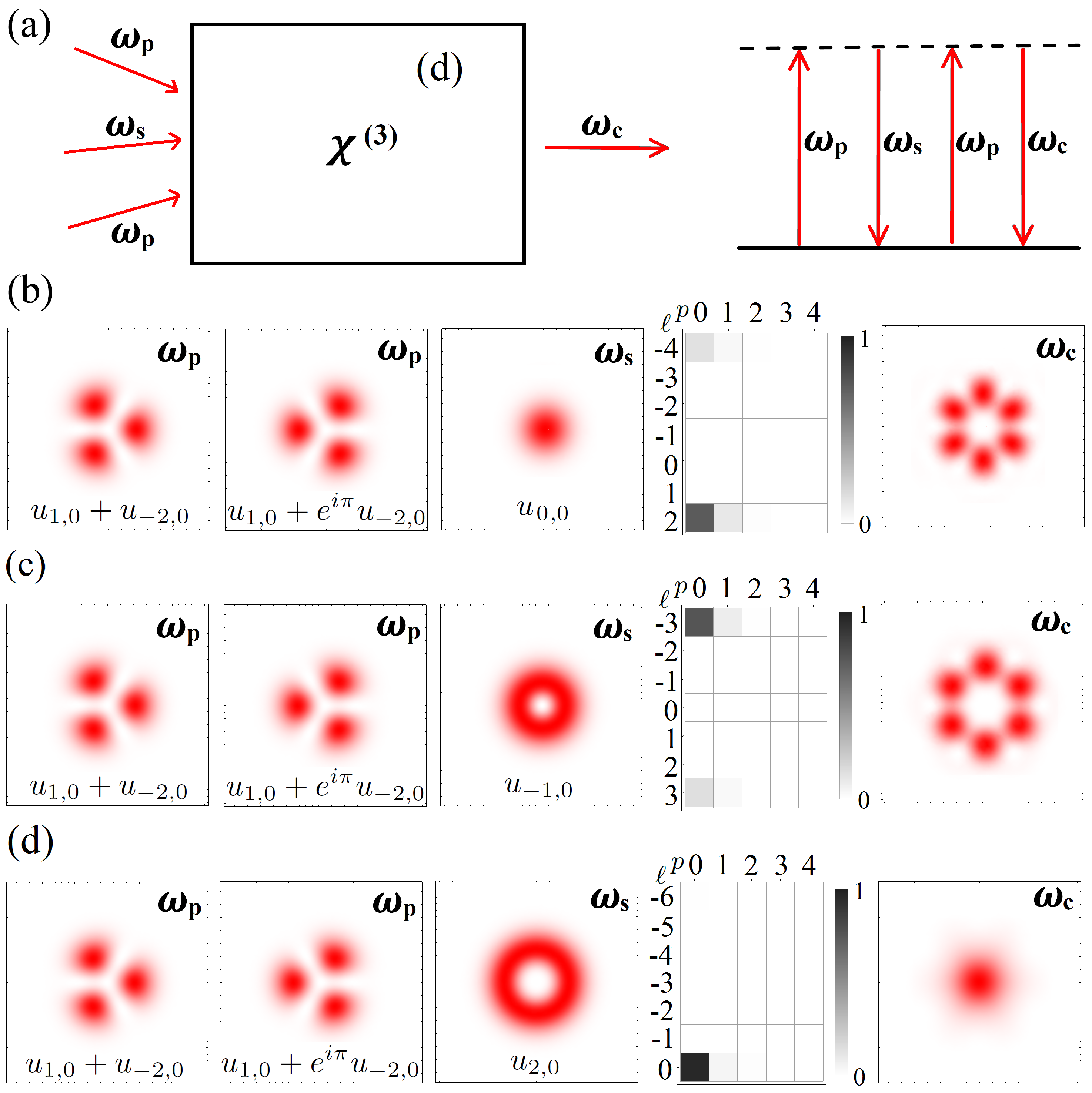}
	\caption{\label{fig:FWM}In (a) we depict the simple degenerate four-wave mixing scheme of our simulation. In (b) through (d) we show the pump-beam profiles, the signal-beam profile, a histogram giving the mode structure of the output (conjugate) beam, and the spatial profile of the output (conjugate) beam respectively. In (b) the pump beams are of the form $u_{1,0}+u_{-2,0}$ and $u_{1,0}+e^{i\pi}u_{-2,0}$ and the signal beam is a $u_{0,0}$ mode. In (c) the pump beams remain the same and the signal beam is a $u_{-1,0}$ mode. In (d) the pumps stay the same but the signal beam is changed to $u_{2,0}$. This is a degenerate scheme and all beams are $650 \, \mathrm{nm}$.
	}
\end{figure}

Next, we present a simulation for degenerate four-wave mixing, in which we demonstrate the addition, subtraction, and cancellation of OAM. 
Two pump beams and a signal interact in a 7 cm long nonlinear cell to create a conjugate beam according to the polarization $\mathcal{P}(\omega_c)^{(3)}=\epsilon_{0} \chi^{(3)} \mathcal{E}(\omega_p)\mathcal{E}(\omega_p)\mathcal{E}(\omega_s)^*$, depicted in Fig.~\ref{fig:FWM}(a). 
We now investigate ways to tailor the response of the conjugate beam. 
In both cases the pump beams remain in the form $u_{1,0}+u_{-2,0}$ and $u_{1,0}+e^{i\pi}u_{-2,0}$, and we choose to vary the signal beam. 
We showed in the first three simulations that, when the pump beams are properly rotated with respect to each other, that cross terms can be canceled, i.e., response at $\ell =-1$ is suppressed.
Therefore, if the seed beam were simply a $u_{0,0}$, then the conjugate beam would respond at $\ell =2$ and $\ell =-4$, as we see in Fig.~\ref{fig:FWM}(b). 
However, suppose we would prefer the field to respond at $\ell =\pm3$, then we would choose the signal to be in a $u_{-1,0}$ mode and effectively add another unit of angular momentum to the conjugate beam, as seen in Fig.~\ref{fig:FWM}(c). 
This approach can be taken to the extreme by choosing the seed to be in a $u_{2,0}$ mode and effectively subtracting two units of angular momentum from the conjugate beam, as seen in Fig.~\ref{fig:FWM}(d). 
In this case, response at $\ell =0$ and $\ell =-6$ are the two modes allowed by OAM conservation. However we see that response at $\ell =-6$ is naturally suppressed due to the poor overlap with the pump and signal beams. 
Thus, we expose one of possibly many ways to completely suppress OAM transfer into unwanted modes.
\section{Conclusion}
In conclusion, we report a general theory for calculating the Laguerre-Gauss mode structure of beams generated during nonlinear interactions. 
We accomplish this by making a first order Born-like approximation to the inhomogeneous paraxial wave equation. 
Therefore, it is akin to the weak scattering problem of the nonlinear Schr$\ddot{\mathrm{o}}$dinger equation, and the solution is consistent with the intuitive picture that the probability of a mode being excited is proportional to the overlap of the modes interacting in the nonlinearity.  
The theory is general in the sense that, it may be implemented for any complete orthonormal set of spatial mode functions. 
We use the theory to simulate orbital angular momentum transfer in several nonlinear optical processes, with an emphasis on analyzing and tailoring the resulting mode structure for optimal performance. 
This includes the processes of orbital angular momentum addition, subtraction, and cancellation in harmonic generation, parametric down conversion, and four-wave mixing.

Although this theory is based on a first order approximation, we show that it can model the mode structure of experimental beam data remarkably well. 
In doing so we are able to show that a naive analysis of the lobe structure in intensity patterns can be very misleading. 
One may be tempted to assume based on lobe structure that OAM transfer is limited to certain pathways.
However, our analysis shows that the rich mode structure resulting from a more complete consideration of all the pathways, actually accounts for symmetries in the resulting beam pattern. 
On a final note, this validation of the theory is assurance that the theory is accurately predicting the classical mode structure, which can in turn be used for second-quantized treatments of these interactions.  

This research was supported by the Air Force Office of Scientific Research grant FA9550-13-1-0098. In addition, RNL, ZX and JDP acknowledge additional support from the Army Research Office, National Science Foundation, and Northrop Grumman Corporation. The authors would also like to thank Sonja Franke-Arnold and her group for sharing the data appearing in Fig.~\ref{fig:FWM2}.

\bibliography{bibliography/bibliography}
\newpage
\appendix
\section{}
\subsection{Green Function Solution Method}
In Sec. II, we introduced our spatial mode propagation equation and constructed an IVP to restate the problem in a compact form. 
The first step in solving our inhomogeneous IVP [Eq.~(\ref{eq:DE2})] is to solve the homogeneous IVP in free space, i.e., for $\wp=0$:
\begin{equation} \label{eq:HOMO_DE}
\begin{split}
\mathrm{DE:}&\;\; \hat{L}\mathcal{E}=0\\
\mathrm{IV:}&\;\;\;\;\;\; \mathcal{E}_{0}.
\end{split}
\end{equation} 
Using a Green function method \cite{Barton_book}, we search for a propagator $K$ defined by
\begin{subequations} \label{eq:PROPAGATOR}
\begin{gather}
\hat{L}\;K(\textbf{r}\;|\; \textbf{r} ') = 0 \quad (z>z')\\
K(\textbf{r}\;|\; \textbf{r} ')=\delta(r-r')\delta(\phi-\phi') \quad (z=z')\\
K \rightarrow 0 \quad (r \rightarrow \infty),
\end{gather}
\end{subequations}
such that when $K$ is known, the homogeneous problem is solved:
\begin{equation}\label{eq:HOMO_SOLN}
\mathcal{E}_{\mathrm{hom}}=\int r'dr'd\phi'\,K(\textbf{r}\;|\; \textbf{r} ')\mathcal{E}_{0}(\textbf{r}')\;|_{z=z'}. 
\end{equation}
The $z=z'$ restraint is a standard condition placed by the boundary conditions which define the propagator, i.e., Eq~(\ref{eq:PROPAGATOR}).  
\begin{proof}
We check our solution Eq.~(\ref{eq:HOMO_SOLN}) by operating the differential operator $\hat{L}$ and find
\begin{equation}
\begin{split}
\hat{L}\mathcal{E}_{\mathrm{hom}}=&\int r'dr'd\phi'\, \big(\hat{L}\;K(\textbf{r}\;|\; \textbf{r} ')\big)\,\mathcal{E}_{0}(\textbf{r} ')\\
=&\int r'dr'd\phi'\, \big( 0 \big)\,\mathcal{E}_{0}(\textbf{r} ')\\
=& 0.
\end{split}
\end{equation}
The second line follows trivially since the propagator $K$ obeys Eq.~(\ref{eq:PROPAGATOR}a). 
Thus we see $\mathcal{E}_{\mathrm{hom}}$ is a solution to Eq.~(\ref{eq:HOMO_DE}). 
\end{proof}

To proceed, we recognize the property in Eq.~(\ref{eq:COMPLETE}) and explicitly define our propagator in terms of the LG modes:
\begin{equation}\label{eq:PROPAGATOR2}
K(r,\phi,z|r',\phi',z') \equiv \sum_{\ell ,p}\,u_{\ell ,p}^{*}(r',\phi',z')u_{\ell ,p}(r,\phi,z). 
\end{equation}
Inserting Eq.~(\ref{eq:PROPAGATOR2}) into Eq.~(\ref{eq:PROPAGATOR}), we verify our choice of the propagator: 
\begin{subequations} \label{eq:CHECK1}
\begin{gather}
\hat{L}\;K = \sum_{\ell ,p}\,u_{\ell ,p}^{*}(\textbf{r}\,')\big(\hat{L}\,u_{\ell ,p}(\textbf{r})\big)= 0 \quad (z>z')\\
K(\textbf{r}\;|\; \textbf{r} ')=\delta(r-r')\delta(\phi-\phi') \quad (z=z')\\
u_{\ell ,p} \rightarrow 0 \ \Longrightarrow \ K \rightarrow 0 \quad (r \rightarrow \infty). 
\end{gather}
\end{subequations}
Therefore, the homogeneous problem is solved, and we can turn our attention back to the inhomogeneous problem Eq.~(\ref{eq:DE2}). 

To solve the inhomogeneous problem, again in free space, we need a Green function $G$ defined by
\begin{subequations} \label{eq:GREEN}
\begin{gather}
\hat{L}\;G(\textbf{r}\;|\; \textbf{r} ') = \delta(r-r')\delta(\phi-\phi')\delta(z-z')\\
G(\textbf{r}\;|\; \textbf{r} ')=0 \quad (z<z')\\
G \rightarrow 0 \quad (r \rightarrow \infty),
\end{gather}
\end{subequations}
such that when $G$ is known, the inhomogeneous problem is solved:
\begin{equation}\label{eq:INHOMO_SOLN}
\mathcal{E}= \int r'dr'd\phi'dz' \, G(\textbf{r} \, | \, \textbf{r} ')\wp(\textbf{r}'). 
\end{equation}
\begin{proof} We check our solution Eq.~(\ref{eq:INHOMO_SOLN}) by operating the differential operator $\hat{L}$ and find
 \begin{equation}
 \begin{split}
\hat{L}\mathcal{E} &= \int r'dr'd\phi'dz' \,\big(\hat{L}\,G(\textbf{r} \, | \, \textbf{r} ')\big) \,  \wp(\textbf{r} ')\\
&= \int r'dr'd\phi'dz'\,\delta(\textbf{r}-\textbf{r} ') \, \wp(\textbf{r} ')\\
&=\wp(\textbf{r}).
\end{split} 
\end{equation}
The second line follows from our definition of the Green function, Eq.~(\ref{eq:GREEN}a), and the third line follows trivially from the definition of the Dirac delta function. Thus we see $\mathcal{E}$ is a solution to Eq.~(\ref{eq:DE2}).
\end{proof} 
Building on the method for solving the homogeneous problem, we explicitly define our Green function $G$ in terms of the LG modes
\begin{equation}\label{eq:GREEN2}
G(r,\phi,z|r',\phi',z') \equiv \Theta(z-z') K(r,\phi,z|r',\phi',z'),
\end{equation}
where $\Theta(z-z')$ is the Heaviside step function. Next, inserting Eq.~(\ref{eq:GREEN2}) into Eq.~(\ref{eq:GREEN}), we  verify our choice of Green function: 
\begin{equation} \label{eq:CHECK2}
\begin{split}
\hat{L}\;G &= (\frac{\partial}{\partial z}-\frac{i}{2k}\nabla^{2}_{\perp})\Theta(z-z') K(r,\phi,z|r',\phi',z')\\
&=K\dfrac{\partial}{\partial z}\Theta(z-z')+\Theta \dfrac{\partial}{\partial z}K-\Theta \dfrac{i}{2k}\nabla^{2}_{\perp}K\\
&=K\delta(z-z')+\Theta(z-z') \, (\dfrac{\partial}{\partial z}K-\dfrac{i}{2k}\nabla^{2}_{\perp}K)\\
&=K\delta(z-z')+\Theta(z-z')\, \hat{L}K\\
&=\delta(r-r')\delta(\phi-\phi')\delta(z-z'),
\end{split}
\end{equation}
and thus we see that Eq.~(\ref{eq:GREEN}a) is satisfied, secondly, Eq.~(\ref{eq:GREEN}b) is satisfied since we know $G|_{z<z'}=0$, by definition of the Heaviside function, and lastly,  Eq.~(\ref{eq:GREEN}c) is satisfied since $r\rightarrow 0 \Longrightarrow u_{l,p}\rightarrow 0 \Longrightarrow G\rightarrow 0$. Therefore, we have derived a valid propagator and Green function and can now write the final solution by combining our two previous solutions Eq.~(\ref{eq:HOMO_SOLN}) and Eq.~(\ref{eq:INHOMO_SOLN}). However, we would first like to simplify the notation in Eq.~(\ref{eq:INHOMO_SOLN}). It is likely that the source $\wp$ doesn't contribute until some position $z'=z_i$, and for $z>z'=z_i$, the Heaviside function reduces to $1$, leaving only the propagator $K$. Furthermore, the source likely only contributes up to some position $z_f$ and thus we may safely modify the $dz$ integral. 
With these final considerations, we can write the complete solution to Eq.~(\ref{eq:DE2}):
\begin{equation}\label{eq:FINAL_SOLN1}
\begin{split}
\mathcal{E}& =  \int r'dr'd\phi'\,K(\textbf{r} \, | \, \textbf{r} ')\mathcal{E}_{0}(\textbf{r} ')\;|_{z=z'}\\
&+ \int_{z_i}^{z_f} dz' \int r'dr'd\phi' \, K(\textbf{r} \, | \, \textbf{r} ')\wp (\textbf{r} ')\\
&= \; \mathcal{E}_{0}\\
&+ \sum_{l,p} u_{l,p}(\textbf{r}) \int_{z_i}^{z_f} dz' \int r'dr'd\phi' \,u_{l,p}^{*}(\textbf{r} ') \wp (\textbf{r} ').
\end{split}
\end{equation}
Although the problem is solved, we can further employ the LG modes to simplify our calculations. Suppose we expand the nonlinear source in terms of the LG modes, i.e.,
\begin{equation}\label{eq:RHO}
\wp(r,\phi,z)=\sum_{l,p}c_{l,p}(z) \, u_{l,p}(r,\phi,z),
\end{equation}
where $c_{l,p}(z)=\int r dr d\phi\,u_{l,p}^*(r,\phi,z)\wp(r,\phi,z)$. 
One should notice that, in contrast to expanding an arbitrary Gaussian beam in terms of the LG modes, the expansion coefficients for a nonlinear source distribution \textit{are not independent of the position $z$}. 
Thus, each infinitesimal slice of the source contributes to the new field $\mathcal{E}$ in the following way. 
We insert Eq.~(\ref{eq:RHO}) into Eq.~(\ref{eq:FINAL_SOLN1}) and invoke the orthogonality of the LG modes one last time to arrive at 
\begin{equation}
\mathcal{E}(\textbf{r})=\mathcal{E}_0(\textbf{r}) + \sum_{l,p}u_{l,p}(\textbf{r})\int_{z_i}^{z_f} dz' c_{l,p}(z').
\end{equation}
This result corresponds to Eq.~(\ref{eq:FINAL_SOLN2}) from the main text.

Lastly, we emphasize the $z$-dependence of the $c_{l,p}$ coefficients, which represent the amplitudes of the new mode structure of the beam.
On one hand, these coefficients allow one to study how the new beam evolves during the interaction in the nonlinear medium.
On the other, in the case that the first Born approximation does not hold, one can use this theory for successive iterations.  
For example, if the pump beam is known to be modified by the nonlinear material, then one can model this by a propagation equation for the pump, and use the solution as a second Born approximation for the new frequency components of the field, i.e.,
\begin{equation}
\wp(\mathcal{E}_0) \rightarrow \wp(\mathcal{E}_1) = \wp \Big( \mathcal{E}_0 + c_{l,p}(z) \, u_{l,p}(\textbf{r}) \Big),
\end{equation}
where in this case $c_{l,p}$ are the result of a propagation equation for the pump.
    
\subsection{Noncollinear Beam Geometry}
\begin{figure}[b]
	\includegraphics[width=1.0\columnwidth]{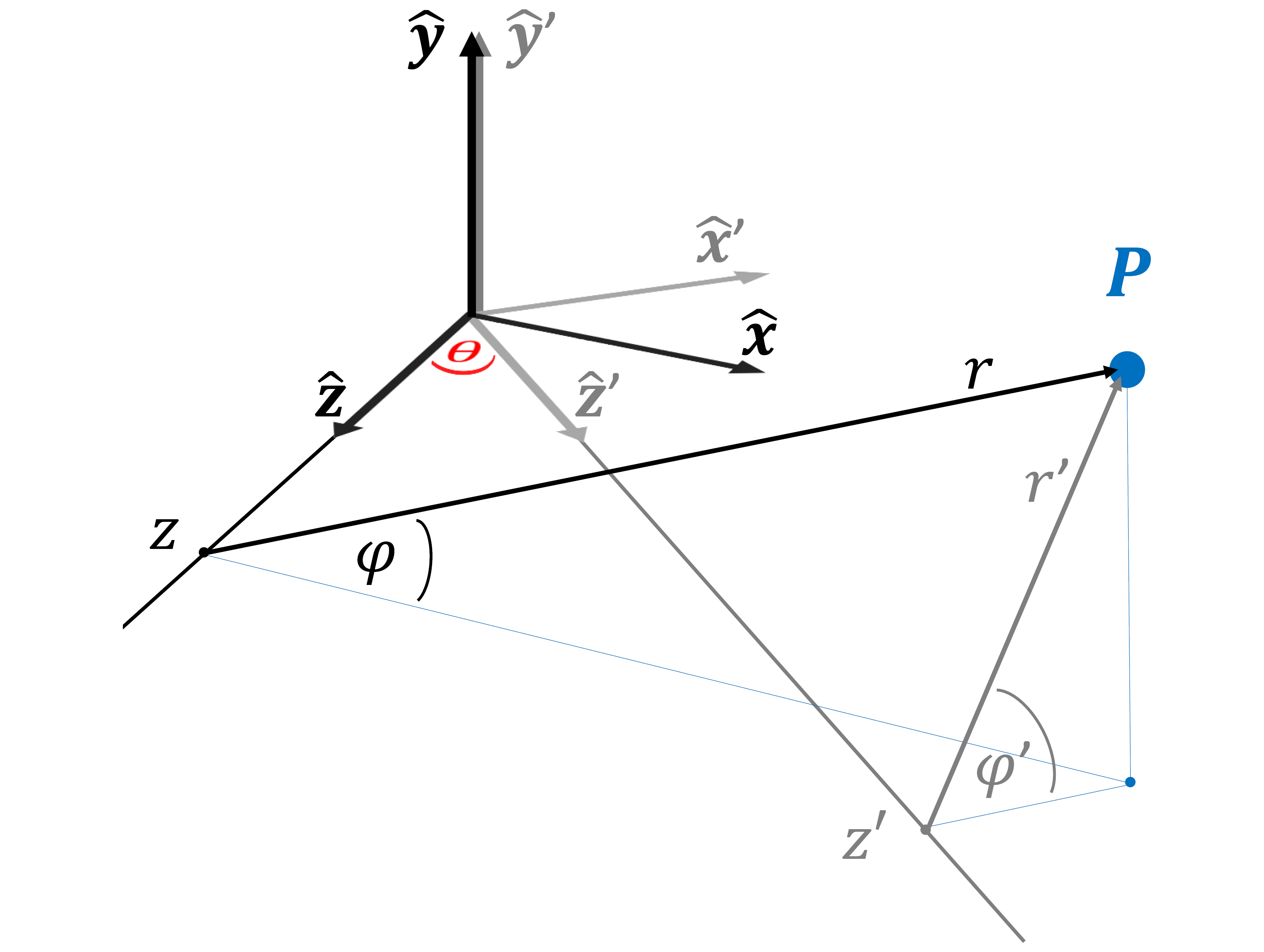}
	\caption{\label{fig:AngledBeams}Geometry of the two-beam coordinate systems, which are separated in the $x\mathrm{-}z$ plane by the angle $\theta$. Any point $P$ can be described by the cylindrical coordinates ($r,\phi,z$) or ($r',\phi',z'$). We associate the generated beam with the un-primed coordinate system, and the input beam with the primed coordinate system; then using geometry we deduce the form of $r',\phi',z'$ in terms of $r,\phi,z,$ and $\theta$. 
	}
\end{figure}

We restrict our problem to only consider beams which are separated by some angle $\theta$, yet lie in the same plane and have focal points which coincide. 
One beam, typically the generated beam, will define the reference frame and the coordinates of the rotated beam will be transformed into this frame (see the geometry in Fig.~\ref{fig:AngledBeams}).
The electric field at $P$ is described by a slice of the reference beam at $z$, and a slice of the tilted beam at $z'$.
One can easily deduce the form of $r',\phi', \mathrm{and} \, z'$, in terms of $r,\phi,z,$ and $\theta$, using this geometry.
However, the transformations make for cumbersome calculations. Therefore, since $\theta$ is small in our simulations, we expand around $\theta$ and keep only the first-order correction for each transformed coordinate:
\begin{equation}
\begin{split}
r'&= r + \theta \times r \,  \sin \phi \, | \cos \phi | \\
\phi ' &= \phi + \theta \times \cos \phi \, | \cos \phi | \\
z' &= z+ \theta \times r \, \cos \phi \, \mathrm{sign}(z) . 
\end{split}
\end{equation}
These corrections made no significant changes in our calculations, for which all beams interact at $\theta \leq 2 $ degrees.

\end{document}